# The role of liquid-liquid phase separation in regulating enzyme activity

Brian G. O'Flynn[1], Tanja Mittag[1*]

Department of Structural Biology, St. Jude Children's Research Hospital, Memphis, TN 38105, USA

* Corresponding author: tanja.mittag@stjude.org

**Graphical abstract**

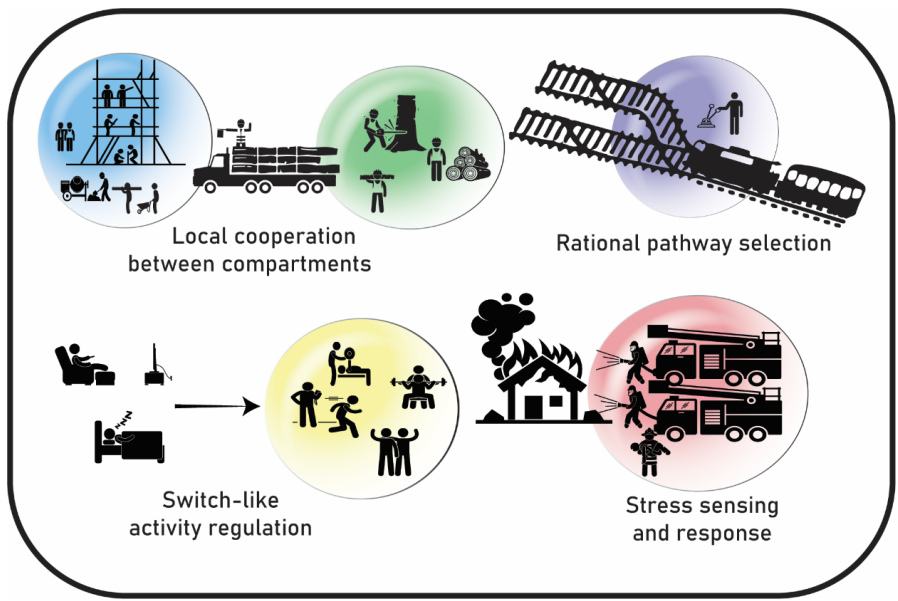

**Abstract**

Liquid-liquid phase separation is now recognized as a common mechanism for regulating enzyme activity in cells. Insights from studies in cells are complemented by *in vitro* studies aimed at developing a better understanding of mechanisms underlying such control. These mechanisms are often based on the influence of LLPS on the physicochemical properties of the enzyme's environment. Biochemical mechanisms underlying such regulation include the potential for concentrating reactants together, tuning reaction rates, and controlling competing metabolic pathways. LLPS is thus a powerful tool with extensive utilities for the cell, e.g. for consolidating cell survival under stress or rerouting metabolic pathways in response to the energy state of the cell. By examining the evidence of how LLPS affects enzyme catalysis, we can begin to understand emerging concepts and expand our understanding of enzyme catalysis in living cells.

**Introduction**

Enzymatic activity within the cell is delicately balanced through several layers of organization, primarily illustrated by the various membrane-bound organelles which operate as compartmentalized hubs of reactivity, each specializing in their own functions and maintaining different solution conditions. Recently, the insight that cells can generate another layer of organization through the process of liquid-liquid phase separation (LLPS) has added a further intriguing layer of complexity to the regulation of enzyme catalysis [1].

LLPS is a dynamic and reversible process through which a group of biomolecules, such as proteins, DNA, or RNA, associate via multivalent interactions into a non-stoichiometric assembly. The result of this process is the formation of at least two distinct phases, i.e. a dense phase enriched with the biomolecules and often taking the form of large droplets, and the surrounding dilute phase, which is depleted of the biomolecules [2-4]. Phase separation only occurs once a system-specific threshold concentration of participating biomolecules is surpassed. This is often referred to as the saturation concentration ($c_{sat}$) and is dictated by the molecules' valence, i.e. the number of interactions that can occur between the responsible biomolecules [5].

Evidence is accumulating that LLPS is critical to such cellular processes as cell signaling [6,7], transcription [8-11] and stress response [12-16]. Dysregulation of LLPS has also been causally linked to amyotrophic lateral sclerosis (ALS), inclusion body myopathy (IBM), and other closely related neurodegenerative diseases [17-19], as well as many forms of cancer [20-22]. As the molecular mechanisms of the ties between LLPS and disease are becoming better understood, we are also gaining deeper insights into the functions of biomolecular condensates.

One of the possible functions of condensates is as reaction crucibles, wherein phase separation functions to concentrate enzymes and substrates together and enhance turnover (Figure 1A) or conversely, sequester enzymes and substrate away from each other and reduce turnover [2,3,5,23]. The marrying of enzyme catalysis and LLPS allows for the modulation of many parameters important for enzymatic activity, including local reactant concentrations, presence or absence of reaction influencers, timespans for such localization, and even solvent conditions to be controlled. Several reviews have touched on this topic previously [1,3,24,25], and this review aims to complement these by illuminating recent *in vivo* discoveries of LLPS functioning to control or alter enzymatic reaction rates. We also aim to discuss the various engineered systems which have recently been implemented to examine the effect of LLPS on enzymatic activity. Finally, we will touch upon the biophysical effects of enzymes in dense phases and discuss under which conditions activities would actually be expected to be enhanced.

**Phase separation regulates enzyme catalysis in cells**

The idea that phase separation can be used to generate protective environments and to control rates of product formation is not a novel one. The process was receiving attention as far back as the beginning of the 20$^{th}$ century, when scientists such as Alexander Oparin were investigating the roles of coacervates as protocells in the origin of life [26-28]. This theory was largely respectfully dismissed until recently, when biological condensates began to be linked to an abundance of biological processes. The main reason for this belated acknowledgement is largely due to lack of experimental evidence for condensates in cells and *in vivo* until recently. The sophistication of tools now in use to study the biochemistry of the cell means we are seeing a wealth of new information arise related to the biological function and mechanism of LLPS, and how it links to enzyme catalysis.

Cellular stresses are frequently seen as a key trigger for LLPS, taking advantage of the reversible nature of condensates to sequester, protect, or coalesce various biological components. Stress granules are the most well-documented example [15,17,29,30], but other unique condensates have been shown to form depending on the source of the stress [12-14,16,31]. Recent discussions have addressed the ability of metabolic enzymes to phase separate under nutrient deprivation [31-33]. Additional works by Fuller, *et al*. [34] Jin, *et al*. [35] and Kohnhorst, *et al*. [36] expanded on our understanding of glycolytic (G) bodies (or glucosomes as they have also been called), i.e., membraneless condensates which form under hypoxic stress to co-localize glycolytic enzymes and enhance the rate of glycolysis. G bodies were identified to alter glycolysis rates by measuring the levels of glycolytic metabolites in wild-type cells and in cells that lacked Snf1p, a conserved AMP-activated protein kinase found to be vital for G body formation [35]. Exclusively hypoxic conditions led to increased levels of upstream metabolites and decreased levels of downstream metabolites in ΔSnf1p cells in comparison to wild-type cells. This is consistent with the idea that G bodies function to enhance glycolytic turnover. This was corroborated with experiments wherein wild-type cells were cultured under hypoxic or normal conditions prior to transfer to new media under exclusively hypoxic conditions. Significantly increased rates of glucose consumption were noted for cells which were initially cultured under hypoxic conditions and had pre-formed G-bodies. Additionally, nearby localization of protein complexes associated with fatty acid synthesis, trehalose-6-phosphate synthesis and protein degradation was noted. The increased rate of glycolysis of G bodies, and subsequent increased adenosine triphosphate (ATP) turnover, appears to provide local energy input for other vital cellular processes to continue, even under hypoxic conditions (Figure 1B). This agreed with demonstrated G body formation near presynaptic sites to meet requirements to sustain synaptic activity [37].

Interestingly, the rate-limiting glycolytic enzymes were found in multienzyme clusters [36]. Parallel to this, a relationship between condensate size and shunting of glycolysis intermediates between energy metabolism and anabolic biosynthetic pathways was demonstrated [36]. Through selective promotion and inhibition of the pentose phosphate pathway or the serine biosynthesis pathway, two anabolic pathways that promote nucleotide and amino acid biosynthesis, respectively, it was determined that cells with medium-sized clusters had a propensity to divert metabolites towards the pentose phosphate pathway. In contrast, cells with larger clusters shunted resources towards the serine biosynthesis pathway. The importance of this is noted by the presence of large-sized clusters in various human cancer cells, and absence in non-cancerous human breast cells. Increased serine biosynthetic rates are considered a hallmark of altered glucose metabolism in cancer [38]. The different functions of different-sized clusters are unlikely to be a direct consequence of their size. It is instead likely that the different sizes of clusters reflect the recruitment of different subsets of metabolic enzymes which direct forked metabolic pathways into a specific direction. Switching between these different states may be regulated via post-translational modifications or other inputs from the energy state of the cell (Figure 1C).

G bodies are ribonucleoprotein granules, formed through multivalent protein-protein and protein-RNA interactions [34]. Glycolytic enzymes co-localize with a modest amount of RNA to G bodies in hypoxic conditions, i.e. when oxygen levels are reduced. In fact, RNA serves as a scaffold and is essential for G body formation (Figure 1D). 10 glycolysis enzymes were found to have RNA binding properties even under normal oxygen levels, and they all bind similar RNA sequences, including RNA sequences found in transcripts of glycolysis enzymes. These observations suggest that in the case of G bodies, RNA scaffolding and recruitment has the additional potential for localized translation of the proteins that make up the G body and facilitate its growth. This may be an example in which LLPS does not only influence enzyme activity but also *vice versa*.

Similar trends have been noted in the purine biosynthesis pathway, in which six enzymes required for *de novo* purine biosynthesis cluster in 'purinosomes' [39-43]. Increased formation of purinosome condensates under hypoxic conditions was demonstrated, similar to G body formation [40]. In fact, treatment with inhibitors of glycolysis and pentose phosphate pathways in cells under hypoxia led to a reduction in purinosome formation pointing to the interconnectivity of these systems. In contrast to G bodies [37], it was determined that there was no up-regulation of enzyme activity within the purinosome under hypoxic conditions. This was attributed to a subsequent downregulation of mitochondrial one-carbon metabolism under hypoxia and provides a caveat to previous reports stating that purinosomes increased *de novo* purine biosynthesis [40-42]. The specific reasons as to why purinosome formation is increased under hypoxic conditions therefore remains to be uncovered.

The regulation of enzymatic function by LLPS is not limited to metabolic enzymes, with numerous examples of LLPS functioning to control translation and deadenylation rates. LLPS of Argonaute2 and TNRC6B, two components of the human miRNA-induced silencing complex (miRISC), was shown to sequester target RNA from the surrounding solution, leading to an increased rate of deadenylation [44]. Similarly, FMRP and CAPRIN1, regulator proteins which have been shown to be involved in mRNA stability and translational repression, were found to phase separate with target RNA, but only after phosphorylation of FMRP. Differential Ser/Thr or Tyr phosphorylation tunes rates of deadenylation and translation [45]. This implies that phase separation does not only lead to the recruitment of RNA substrates together with their respective enzymes to enhance activity, but that it can lead to the up- or down-regulation of one reaction pathway over another, simply by tuning the physical characteristics of the condensate via signaling or metabolic inputs (Figure 1C), similar to phenomena observed with G bodies [36]. In contrast, phase separation of cytoplasmic poly(A)-binding protein (PAB1) was found to regulate translation in a very different manner [46]. PAB1 binds to RNA and acts as a translational repressor. Upon phase separation the protein is sequestered away from RNA allowing for its subsequent translation (Figure 1D). These findings give further credence to the diverse but conceptually similar roles LLPS can play in enzyme catalysis.

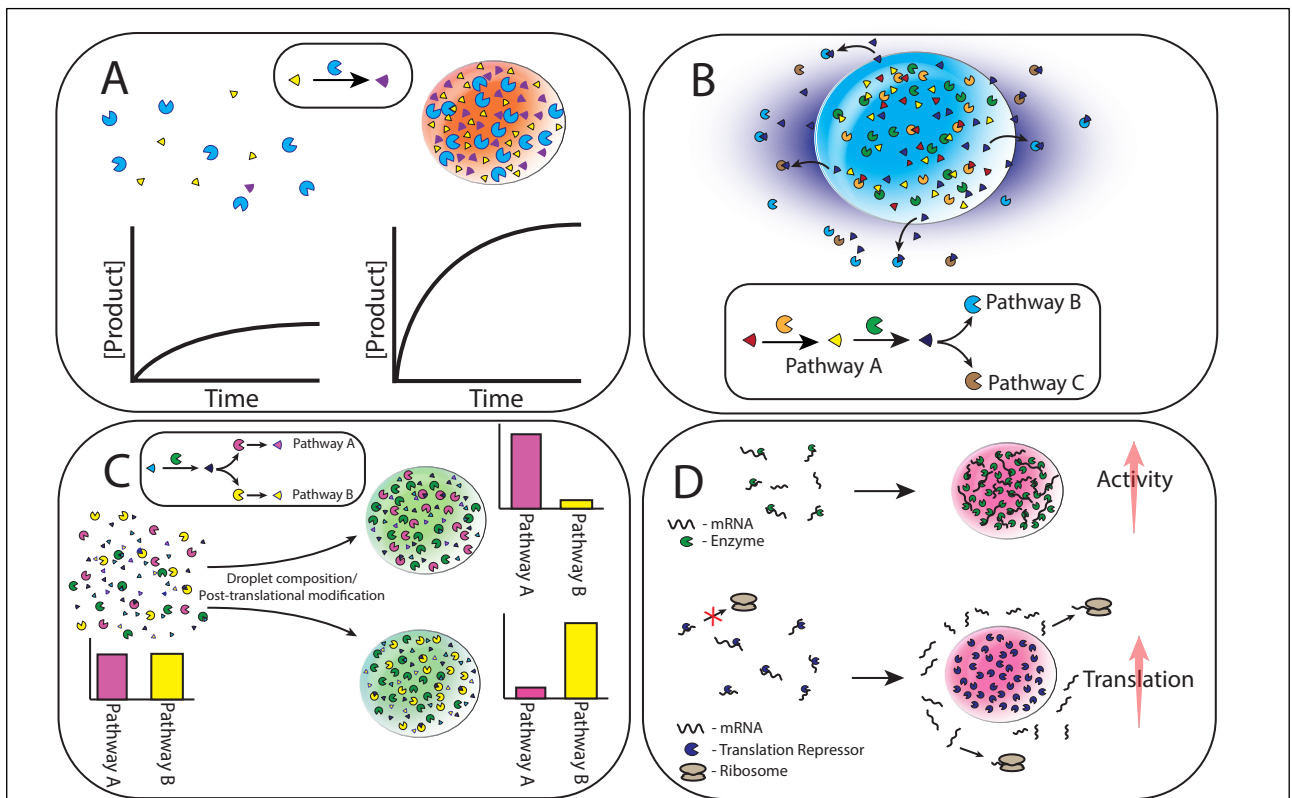

**Figure 1: Cells use LLPS to control enzyme activity. A.** LLPS can increase the local concentration of reactants, thus increasing rates of product formation. **B.** LLPS of enzymes within one particular pathway (pathway A) can lead to increased production of a reagent used in other pathways (pathways B and C). The localized production of the reagent results in localization of these enzymes around the condensate (or even their partitioning into the condensate) [35]. **C.** Selective co-compartmentalization of certain enzymes in a forked metabolic pathway can divert metabolites toward one pathway. Switching between alternate compartmentalization states may be regulated by post-translational modification of phase-separating proteins, e.g. as a result of the energy state of the cell. **D.** RNA-binding proteins can phase separate with mRNA and control its translation and degradation, locally and globally.

Taken together, these data demonstrate the multifunctional potential of biomolecular condensates which can be summarized as: 1. Enhancement of activity to negate effects of stress and/or meet local physiological demands (Figure 1A). 2. Selective transfer of substrates between enzymes via the recruitment of enzymes into the vicinity of enzymatically active condensates (Figure 1B). 3. Redirection of metabolites between multiple diverging pathways based on cellular demand by selective compartmentalization (Figure 1C). 4. mRNA binding capabilities to protect cognate mRNA, regulate translation efficiency, and facilitate LLPS (Figure 1D). Of interest, Fuller *et al*. noted the identification of 69 novel mRNA binding proteins, demonstrating the potentially ubiquitous role of LLPS in regulation of many metabolic pathways [34].

**Investigations into the effect of LLPS on enzyme activity in *de novo* designed systems**

The pursuit of understanding how phase separation can function *in vivo* to mediate or regulate enzyme catalysis is complemented by *in vitro* investigations. Designed systems can push the limits of phase separation and enzyme catalysis and answer questions relating to general mechanisms and functional benefits of phase separation on enzyme catalysis and its limits. The broad range of design options available to study the relationship between phase separation and enzyme catalysis means that much of our understanding of how these processes influence each other comes from relatively straightforward engineered systems. These systems often incorporate components from the biological world to tune phase separation abilities. By linking so-called 'scaffold' molecules, those that control the propensity to phase separate, with an enzyme with a measurable readout, a system is developed which can offer insights into how phase separation affects reaction rates [47]. Adenylate kinase coupled with low complexity domains from Dbp1 and Laf1, a pair of DEAD-box proteins which have been shown to phase separate *in vivo*, resulted in a significant increase (60- to 100-fold) in enzyme concentration within droplets as compared to the starting concentration and a five-fold increase in initial velocity of reaction when compared to a homogenous solution [48,49]. This increase was attributed to the increased local concentration of reactants but was also notably limited by the increase in viscosity (by 25- to 200-fold), leading to a proposed limitation in turnover. In fact, viscosity has been proposed to limit activity increases in condensates in several cases [2,22] (Figure 2A).

RNA-scaffolded condensates have been implicated in the development of prebiotic early life [50]. To test the viability of this proposal, condensates of the polycation polydiallyldimethylammonium chloride (PDAC) and a polyA-RNA molecule were formed and shown to alter the rates of ribozymes [51]. The electrostatic interactions that drive formation of RNA condensates were attributed for a 60-fold decrease in activity for the hammerhead ribozyme by disrupting native structure [52]. However, when bulk ribozyme concentrations were kept below the dissociation constant ($K_d$), activity was increased up to 10-fold in the dense phase when compared to the dilute phase due to a sharp increase in local concentration upon coacervation. This proposes an interesting mechanism of reaction rate control based around the saturation concentration of an enzyme and the $K_d$ or $K_M$ of the enzyme with its substrate.

Artificial IDPs (A-IDPs) with controllable LLPS behavior were shown to recruit small molecules and proteins into condensates and so it was hypothesized that they could be used in the design of enzymatically active condensates [53]. To ensure that the expression of a bulky enzyme attached to the A-IDPs would not affect LLPS, a split version of β-galactosidase was utilized, wherein the active enzyme was only formed when two components of β-galactosidase were recombined. This also had the advantage that the enzyme was only active when the A-IDPs formed complexes. The versatility of the A-IDP system allowed the researchers to investigate the effects of increased A-IDP chain length and aliphatic/aromatic content on kinetic rates. It was found that with larger A-IDP chains comes an increased enzymatic efficiency – 2.5 and 7.5 times greater for a 40mer and 80mer respectively when compared to a control. Measurable kinetic constants, namely $k_{cat}$ and $K_M$, meant that the increased rates could be attributed to a higher turnover rate ($k_{cat}$) rather than an increased propensity to form an enzyme-substrate complex ($K_M$). Interestingly, no changes in rates were observed when aromatic content of the A-IDP was altered suggesting that the increase in $k_{cat}$ is decoupled from saturation concentration.

Phase separation has the ability to increase enzyme activity by more than expected by mass action [54]. An engineered system that used multivalent SH3 domain-containing proteins and proline-rich motif-containing proteins as scaffolds and recruited enzyme and substrate of a SUMOylation pathway, showed 36-fold increased activity of the dense over the dilute phase. Unexpectedly, the total reaction volume had 7-fold higher activity in the presence of phase separation than in the absence, which stemmed from excess activity in the dense phase beyond what was expected by mass action. This effect was only evident for specific scaffold variants and was accompanied by a lower $K_M$ value in the dense phase. It was likely caused by the enhanced proximity of enzyme and substrate induced by sterically favorable recruitment to the scaffold. LLPS can thus not only enhance enzyme activity by mass action but also by preorganization of enzyme/substrate complexes [54].

The ability of condensates to facilitate enzymatic catalysis is not limited to single-step reactions. By combining the phase separating ability of a trio of synaptic proteins, GKAP, Shank, and Homer, with high-affinity interacting peptides, RIDD and RIAD, a phase-separating system with the ability to selectively recruit 'guest' proteins was engineered [55]. This system was used to examine the effect of phase separation on the multienzyme menaquinone biosynthesis and terpene biosynthesis pathways. While an increase in the rate of final product formation was noted for both systems when components of the systems were recruited to the condensate, it is interesting to note that each had opposite effects when not recruited. Of the three enzymes in the menaquinone biosynthesis pathway, MenD, MenF, and MenH, only MenH could be tagged with RIDD and recruited to the condensate. It was found that both tagged and untagged systems had the same reaction rate, regardless of active recruitment, in comparison to a homogenous solution with no phase separation and a lower rate. In contrast, the recruitment of both enzymes in the terpene biosynthesis pathway, Idi and IspA, had over a 50% increase in activity over the untagged and homogenous controls. The lack of activity enhancement upon recruitment of MenH likely has to do with MenD, not MenH, being the rate-limiting enzyme in the menaquinone biosynthesis pathway [56]. Low levels of enrichment of untagged MenD likely negated any effect of actively recruited MenH. An increase in rate of product formation may only be of use to a cell if the recruited enzyme catalyzes a rate-limiting step in a multienzyme pathway.

Utilization of scaffold proteins derived from biological systems offers the ease of genetically encoded constructs and a biologically relevant outlook on the effect of LLPS on enzyme catalysis. Incorporating or substituting scaffold proteins with more artificial components, such as nanoparticles, polymers, or crowding agents, can offer readily tunable phase separation *in vitro* and the exploration of enzyme activity in synthetic protocells [52,57-62]. These are complementary and valuable resources that can provide a robust viewpoint on how enzyme catalysis is influenced by the effects of LLPS. A system composed of three separate liquid phases was used to examine the effect of compartmentalization on a cascade of enzymatic activities [63]. Horseradish peroxidase (HRP) along with a chromogenic substrate was encapsulated within an ATP/PDAC coacervate. The surrounding dextran phase contained glucose oxidase. Surrounding this again was a PEG phase rich in glucose. Oxidation of glucose and the concurrent reduction of oxygen to $H_2O_2$ allowed for the subsequent oxidation of the co-compartmentalized chromogenic substrate by HRP. It was found that compartmentalization of HRP with its substrate was necessary, as separate compartmentalization resulted in a significant reduction in activity. Additionally, multiple ATP/PDAC coacervates containing HRP with different substrates demonstrated the potential for multiple separate reactions to occur simultaneously using reactants from one upstream pathway [63]. These results demonstrate the power of compartmentalization to increase functionality.

It has been speculated that LLPS can function as an on/off switch for enzyme activity [5]. This could be the case if enzyme and substrates are too dilute for turnover in the absence of phase separation but sufficiently concentrated in the dense phase. Given the theoretical infinite cooperativity of phase separation, an infinitesimal change in enzyme or substrate concentration would have the potential to suddenly lead to turnover (Figure 2B). Switches are difficult to engineer in biology and often rely on cascades of kinases to sharpen a response. The physical properties of phase separation offer readily available switch-like behavior, and this may well be the reason why phase separation is so ubiquitous in cells. The potential of phase separation to regulate enzyme activity in a switch-like manner has now been shown experimentally. Formate dehydrogenase was co-encapsulated within synthetic vesicles containing poly-lysine components and either carboxymethyl-dextran or ATP [60]. By dropping the pH of the vesicle below the $pK_a$ of poly-lysine (pH 10.5), complex coacervation was induced through protonation of poly-lysine and facilitation of electrostatic interactions between condensate components. Formate dehydrogenase activity is low at low protein concentrations. When encapsulated, formate dehydrogenase activity was "switched on", and the rate of conversion of $NAD^+$ to NADH was increased, in agreement with effects expected for an increase in the local concentration of the enzyme. Following return to pH 11, the system returned to a homogenous mixture, and NADH production was switched off. A magnetic nanoparticle coated with LCDs from Laf1 conjugated to adenylate kinase also allowed for phase separation in response to changes in pH and ionic strength, and morphology alteration in response to a magnetic field [57]. Enzyme catalysis was again only observed under conditions promoting phase separation.

Phase separation does not only have the potential to influence enzyme activity; in fact, phase-separating enzyme systems have been engineered that reveal the potential of enzymatic reactions to alter the driving force for phase separation. Complex coacervates formed from positively charged peptides and RNA are readily dissolvable by phosphorylation of the peptides and reform upon their dephosphorylation [64]. This system is hence regulatable by opposing kinase and phosphatase activities and could be coupled to another enzyme activity that is only active once the condensate is formed. One could even envision a system that periodically cycles between on- and off-states of enzyme activity (Figure 2B).

The importance of artificial systems for our understanding of the impact of LLPS on enzyme activity lies in the ability to experimentally tune the driving force for phase separation. This allows for intricate balances between enzyme rates, substrate affinity, multi-enzyme reaction bottlenecks and enzyme activation to be studied in the context of condensate saturation concentrations, viscosity, material properties, and stability under varying pH and ionic strengths. The continued investigation into such relationships is of paramount importance to the field, to understand the limits and benefits of regulating enzyme activity via LLPS.

**Implications of LLPS on enzyme activity and how a cell can take advantage of this**

Accumulating evidence indicates that cells use LLPS as a method to increase local concentration of reactants. The dynamic nature of condensates offers a wealth of opportunity to fine-tune reactions, function as selective metabolic crossroads, and act as on/off switches to regulate enzyme activity depending on cellular demand (Figure 2C). The mechanistic basis of these effects likely stem from the unique environment that phase-separated condensates offer. Localizing a particular enzyme within a condensate has the power to increase its concentration; by how much is dependent on the system and the conditions. Some phase-separating molecules differ in dilute and dense phase concentration by a thousand-fold under certain conditions [65]; others experience relatively small increases in concentration, particularly in condensates with several components [29]. Each system can be balanced close to a critical point, where the phase-separating components will be hardly enriched in the dense phase. We expect that the partition coefficient of enzymes into condensates is tightly regulated and under evolutionary pressure, but this has not been explored yet.

Not only are enzymes concentrated in condensates, but reactants, cofactors, and products can also be concentrated. Macromolecular substrates can act as scaffolds, which naturally results in their strong recruitment. Alternatively, they may be clients that partition into the dense phase via the same interactions that drive phase separation. Even small molecules can be significantly enriched in dense phases via interactions with macromolecular components [66]. The affinity of substrate and enzyme may be sufficient to achieve such a localization, especially if the substrate is abundant. Importantly, condensate components localize dynamically, can diffuse in and out and create a highly active, yet crowded environment. The effect that such crowding has on enzymatic reactions is important to comprehend, as it is directly tied to how a cell utilizes LLPS to control enzyme activity.

Enzyme crowding has been discussed in detail previously [25,67-70]. However, recent studies from Trylska and coworkers shed new light on the effects of crowding on the activity of enzymes [71-73] (Figure 2C). They showed that the activity of HIV-1 protease was progressively suppressed by the presence of increasing concentrations of polyethylene glycol (PEG) [71]. This effect was also observed in subsequent experiments involving the hepatitis C virus protease NS3/4A [73]. Molecular dynamics simulations showed PEG-protein interactions predominantly involving side-chain residues, but with no particular obstructions of the active site of NS3/4A. This resulted in a decrease in $k_{cat}$, consistent with reduced rate of product formation and release, and pointed to slowing of catalytically important dynamics in the enzyme. Interestingly, Ficoll, a crowder which has been shown to be more inert than PEG, had the opposite effect on activity, with almost an 8-fold increase in rate of activity, likely through excluded volume effects. In addition, Ficoll crowding was found to reduce the activity of Telaprevir, a potent inhibitor of NS3/4A protease. When bovine serum albumin (BSA) was used as the crowder, a different effect was seen again, with a sharp increase in $K_M$ observed. This is consistent with crowding affecting the affinity the substrate has for the enzyme rather than affecting turnover of product [73] (Figure 2C).

It is apparent that not only does the biophysical phenomenon of crowding affect enzyme activity, but the biochemical interactions between the molecules being crowded are of significant importance to how the enzyme activity is affected. This points to the importance of the unique solvent properties found within condensates. It has been proposed that the solvent environment in condensates may reflect conditions closer to an organic solvent than water with substantially decreased dielectric constant. [48,74,75]. This was based on the observation that double-stranded DNA was melted upon partitioning into condensates of the DDX4 low-complexity domain [74]. Akin to a chemist selecting the correct solvent for an organic synthesis experiment, the cell may be able to control reaction rates by altering the local solvent properties (Figure 2A).

Phase separation also allows for the specific recruitment or exclusion of reactants. Dextranase had different kinetic profiles whether partitioned within a phase separated droplet of dextran, or compartmentalized adjacent to the droplet [63]. Rates were shown to be slower when dextranase was found in the same droplet as dextran whereas when separately compartmentalized, dextranase was more efficient. This was hypothesized to be due to substrate inhibition, but is arguably better characterized as product inhibition, as the product of the hydrolysis of dextran is just a smaller molecule of dextran. When a product has a similar structure to the substrate, it can reversibly bind to the free enzyme, inhibiting activity [76]. In this case, dextranase has simply been shown to have reduced activity as dextran size decreases, in agreement with product inhibition [77]. Nevertheless, the results propose an interesting rationale for compartmentalizing components in a reaction separately rather than partitioning all together.

The mechanism behind such proximity-based activity increase is likely also tied to intrinsic biophysical properties of the condensate. Saleh *et al*. investigated circumstances where the phase-separated component consists of the substrate (DNA) whereas the surrounding dilute phase contains the enzyme (a restriction enzyme – SmaI) free in solution [78]. Enzymatic activity, and thus degradation of the droplet, was heightened by those enzymes which penetrated within the droplet. Penetration was most efficient for conditions balanced at the phase boundary and led to the formation of dilute phase vacuoles within the condensate. The modulation of enzyme and/or substrate diffusion across phases is likely another way cells can regulate kinetic rates of enzymes through LLPS.

These findings imply that cells can feasibly control reaction rates simply by crowding an enzyme with a component that either encourages enzyme-substrate interaction, offers preferred environmental conditions, controls quench depth or slows formation and release of a product (Figure 2).

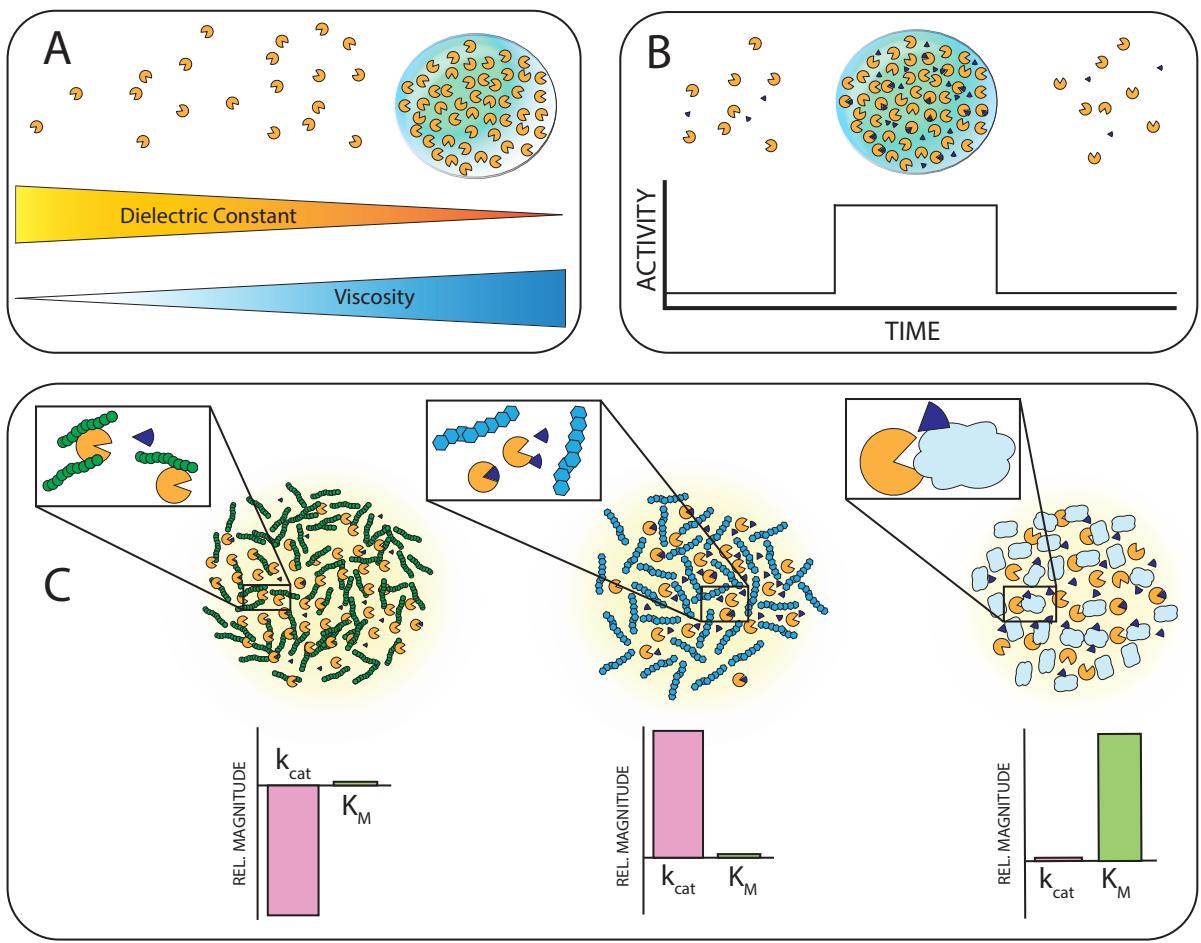

**Figure 2: Biophysical and biochemical effects of phase separation can function to regulate enzyme activity. A.** LLPS results in a unique local environment that differs from the surrounding dilute phase. The solvent properties of condensates can resemble organic solvents, allowing tuning of reaction conditions by LLPS. The resulting increase in viscosity has been shown to reduce enzyme activity due to reduced mobility of reagents. This implies a balance between contrasting effects to realize optimal reaction conditions within condensates [2,22,74,75]. **B.** The high cooperativity of LLPS means that enzymatic reactions can be switched on or off. By balancing the concentration of enzyme at the threshold of phase separation (likely through translational regulation), activity can be rapidly switched on by a small increase in local concentration, and off again by a subsequent small decrease depending on cellular needs [5,14,57,60]. **C.** Crowding within a condensate can result in a variety of effects on kinetic activity. Interactions between components in a condensate can result in a reduced rate of product turnover by freezing enzyme dynamics critical to turnover. Non-interacting components can reduce the available space through their excluded volume, concentrating enzymes together with their substrates, resulting in increased turnover. Finally, components which form weak, nonspecific interactions with the substrate and reduce accessibility to the enzyme active site can result in an apparent reduction in enzyme-substrate affinity [73].

**Conclusion**

A wealth of evidence demonstrating the tight interconnectivity of liquid-liquid phase separation and enzyme activity regulation means much effort is now being extended to fully understanding this link and its implications. Within cells, the delicately balanced parameters that define a biological condensate have been shown to control enzymatic rates, directing reactions towards pathways based on cellular demand, and switching enzyme activity on and off. The emergence of next-generation microscopy techniques, biomolecular engineering capabilities, and advances in enzyme activity quantification offers unparalleled opportunities for the understanding of the extent of this interplay in the future [69,79]. *In vitro* studies will help characterize the mechanisms underlying cellular phenomena, link them to intrinsic properties of biomolecules and define the potential of the process in cell regulation.

Intrinsic properties of enzymes are also being seen in a whole new light. The kinetic constants of $K_M$, $V_{max}$, and $k_{cat}$ have been shown to be strongly dependent on local environment. Thus, LLPS can potentially function to increase or decrease affinity between substrate and enzyme, reduce catalytic mobility of an enzyme, increase the encounter probability of substrate and enzyme, modulate product release, or have other potential effects. Kinetic constants are measurable values, and as such, can allow us to quantify the extent to which LLPS affects enzyme catalysis, understand the mechanism of such an effect, and examine how seemingly unrelated values such as $K_M$ and $c_{sat}$ could, in fact, be linked. It should be noted, however, that such kinetic measurements are tied to control of not just substrate concentration, but also enzyme concentration. Much of the work discussed herein has characterized 'enzyme activity' using initial velocities, which offers limited insight into the true effect of LLPS on an enzyme. Some works that have elucidated kinetic constants are limited in control over local substrate and/or enzyme concentration within condensates. While large leaps have been taken to investigate the link between LLPS and enzyme activity, there is still much to unearth.

**Acknowledgments**

T.M. acknowledges funding by NIH grant R01GM112846, by the St. Jude Children's Research Hospital Research Collaborative on Membrane-less Organelles in Health and Disease, and by the American Lebanese Syrian Associated Charities. The content is solely the responsibility of the authors and does not necessarily represent the official views of the National Institutes of Health.

**Conflict of Interest**

T.M. is a consultant for Faze Medicines. This affiliation has not influenced the scientific content of this review.

**References**

1. Nakashima KK, Vibhute MA, Spruijt E: **Biomolecular Chemistry in Liquid Phase Separated Compartments**. *Frontiers in Molecular Biosciences* 2019, **6**:21.

2. Banani SF, Lee HO, Hyman AA, Rosen MK: **Biomolecular condensates: organizers of cellular biochemistry**. *Nat Rev Mol Cell Biol* 2017, **18**:285-298.

3. Shin Y, Brangwynne CP: **Liquid phase condensation in cell physiology and disease**. *Science* 2017, **357**.

4. Boeynaems S, Alberti S, Fawzi NL, Mittag T, Polymenidou M, Rousseau F, Schymkowitz J, Shorter J, Wolozin B, Van Den Bosch L, et al.: **Protein Phase Separation: A New Phase in Cell Biology**. *Trends Cell Biol* 2018, **28**:420-435.

5. Li P, Banjade S, Cheng H-C, Kim S, Chen B, Guo L, Llaguno M, Hollingsworth JV, King DS, Banani SF, et al.: **Phase transitions in the assembly of multivalent signalling proteins**. *Nature* 2012, **483**:336-340.

6. Su X, Ditlev JA, Hui E, Xing W, Banjade S, Okrut J, King DS, Taunton J, Rosen MK, Vale RD: **Phase separation of signaling molecules promotes T cell receptor signal transduction**. *Science* 2016, **352**:595-599.

7. Case LB, Zhang X, Ditlev JA, Rosen MK: **Stoichiometry controls activity of phase-separated clusters of actin signaling proteins**. *Science* 2019, **363**:1093-1097.

8. Berry J, Weber SC, Vaidya N, Haataja M, Brangwynne CP: **RNA transcription modulates phase transition-driven nuclear body assembly**. *Proc Natl Acad Sci U S A* 2015, **112**:E5237-5245.

9. Sabari BR, Dall'Agnese A, Boija A, Klein IA, Coffey EL, Shrinivas K, Abraham BJ, Hannett NM, Zamudio AV, Manteiga JC, et al.: **Coactivator condensation at super-enhancers links phase separation and gene control**. *Science* 2018, **361**.

10. Boija A, Klein IA, Sabari BR, Dall'Agnese A, Coffey EL, Zamudio AV, Li CH, Shrinivas K, Manteiga JC, Hannett NM, et al.: **Transcription Factors Activate Genes through the Phase-Separation Capacity of Their Activation Domains**. *Cell* 2018, **175**:1842-1855 e1816.

11. Cai D, Feliciano D, Dong P, Flores E, Gruebele M, Porat-Shliom N, Sukenik S, Liu Z, Lippincott-Schwartz J: **Phase separation of YAP reorganizes genome topology for long-term YAP target gene expression**. *Nat Cell Biol* 2019, **21**:1578-1589.

12. Riback JA, Katanski CD, Kear-Scott JL, Pilipenko EV, Rojek AE, Sosnick TR, Drummond DA: **Stress-Triggered Phase Separation Is an Adaptive, Evolutionarily Tuned Response**. *Cell* 2017, **168**:1028-1040 e1019.

13. Franzmann TM, Jahnel M, Pozniakovsky A, Mahamid J, Holehouse AS, Nuske E, Richter D, Baumeister W, Grill SW, Pappu RV, et al.: **Phase separation of a yeast prion protein promotes cellular fitness**. *Science* 2018, **359**.

14. Iserman C, Desroches Altamirano C, Jegers C, Friedrich U, Zarin T, Fritsch AW, Mittasch M, Domingues A, Hersemann L, Jahnel M, et al.: **Condensation of Ded1p Promotes a Translational Switch from Housekeeping to Stress Protein Production**. *Cell* 2020, **181**:818-831 e819.

15. Yang P, Mathieu C, Kolaitis RM, Zhang P, Messing J, Yurtsever U, Yang Z, Wu J, Li Y, Pan Q, et al.: **G3BP1 Is a Tunable Switch that Triggers Phase Separation to Assemble Stress Granules**. *Cell* 2020, **181**:325-345 e328.

16. Yasuda S, Tsuchiya H, Kaiho A, Guo Q, Ikeuchi K, Endo A, Arai N, Ohtake F, Murata S, Inada T, et al.: **Stress- and ubiquitylation-dependent phase separation of the proteasome**. *Nature* 2020, **578**:296-300.
14